# Tradeoffs in Biconic Intake Aerodynamic Design Optimization with Sub-optimal Oswatitsch Solutions

J. P. S. Sandhu,[1] M. Bhardwaj,[2] N. Ananthkrishnan,[3] and A. Sharma[4]
*Yanxiki Tech, 152 Clover Parkview, Koregaon Park, Pune, 411001, India*

J. W. Park[5] and I. S. Park[6]
*Agency for Defense Development, Daejeon, 305-600, South Korea*

**The notion of sub-optimal Oswatitsch solutions is introduced in order to systematically conduct a tradeoff between total pressure recovery (TPR) and intake drag coefficient (CDi) for supersonic intakes. It is shown that the Oswatitsch-optimal TPR for a biconic intake may be enhanced by adding a conical flare which modifies the terminal normal shock into a novel Lambda shock structure. The optimization problem is formulated along the lines of Axiomatic Design Theory with the conical angle pair and the cowl fineness ratio as the two design parameters. Reynolds-averaged Navier-Stokes (RANS) simulations are performed to iteratively arrive at the optimal solutions, with and without an intake length constraint, for a fixed value of the intake mass flow rate. The results are used to generate the Pareto front in the space of the objective functions, which yields the set of solutions between which TPR and Cdi may be traded off for one another. Additionally, an off-Oswatitsch solution, where only the second cone angle is altered from its optimal Oswatitsch value, is obtained and is compared with the sub-optimal Oswatitsch solutions that form the Pareto front.**

---

[1] Senior R&D Engineer, Member AIAA; jatinder@yanxiki.com

[2] Senior R&D Engineer; megha@yanxiki.com

[3] Technical Consultant, Associate Fellow AIAA; akn.korea.19@gmail.com, akn@aero.iitb.ac.in

[4] Principal Engineer; anurag@yanxiki.com

[5] Principal Researcher, Propulsion Division

[6] Principal Researcher, Propulsion Division

## Nomenclature

| | | |
|---|---|---|
| $Cd_i$ | = | intake pressure drag coefficient |
| $C_p$ | = | specific heat at constant pressure |
| $D$ | = | diameter of base of intake |
| $h_c$ | = | diameter of capture stream-tube |
| $L$ | = | length of imaginary fore-body |
| $M_\infty$ | = | free-stream Mach number |
| $P$ | = | total pressure |
| $p$ | = | static pressure |
| $R$ | = | specific gas constant |
| $T$ | = | static temperature |
| $x$ | = | distance measured along a streamline |
| $\gamma$ | = | specific heat capacity ratio |
| $\delta_1, \delta_2$ | = | cone half-angles with respect to the reference axis |
| $\delta_3$ | = | conical flare angle |

## I. Introduction

Axisymmetric conical or semi-conical intakes are used on a variety of aerospace vehicles for delivering free-stream air to the engine [1]. However, ensuring that the intake provides the correct air mass flow rate under all flight conditions is always a challenge. Any mismatch between the combustor air demand and the intake air supply may cause the combustor to flame out or the intake to get unstarted [2, 3]. Additionally, one requires the losses in the intake flow to be kept to a minimum and the flow distortion to be as low as possible. The loss is usually evaluated in terms of the total pressure recovery (TPR), that is, the ratio of total pressure at the intake exit station to the free stream total pressure. Therefore, optimizing the supersonic intake to maximize the total pressure recovery for a given air mass flow rate is an important design objective [4]. Several authors [5 - 11] have over the years used a variety of analytical, empirical, and numerical methods to analyze or optimize supersonic intakes for maximum TPR. However, another equally important objective in supersonic intake design is to minimize the additional drag incurred due to the intake. The intake drag comprises of two components: additive drag due to the capture stream-tube, and pressure and skin-friction drag arising from the external cowl surface [12]. Unfortunately, the requirements of maximizing intake TPR and minimizing intake drag are usually at odds with each other. Thus, a multi-objective design optimization of a supersonic intake typically calls for a tradeoff between TPR and intake drag [13, 14].

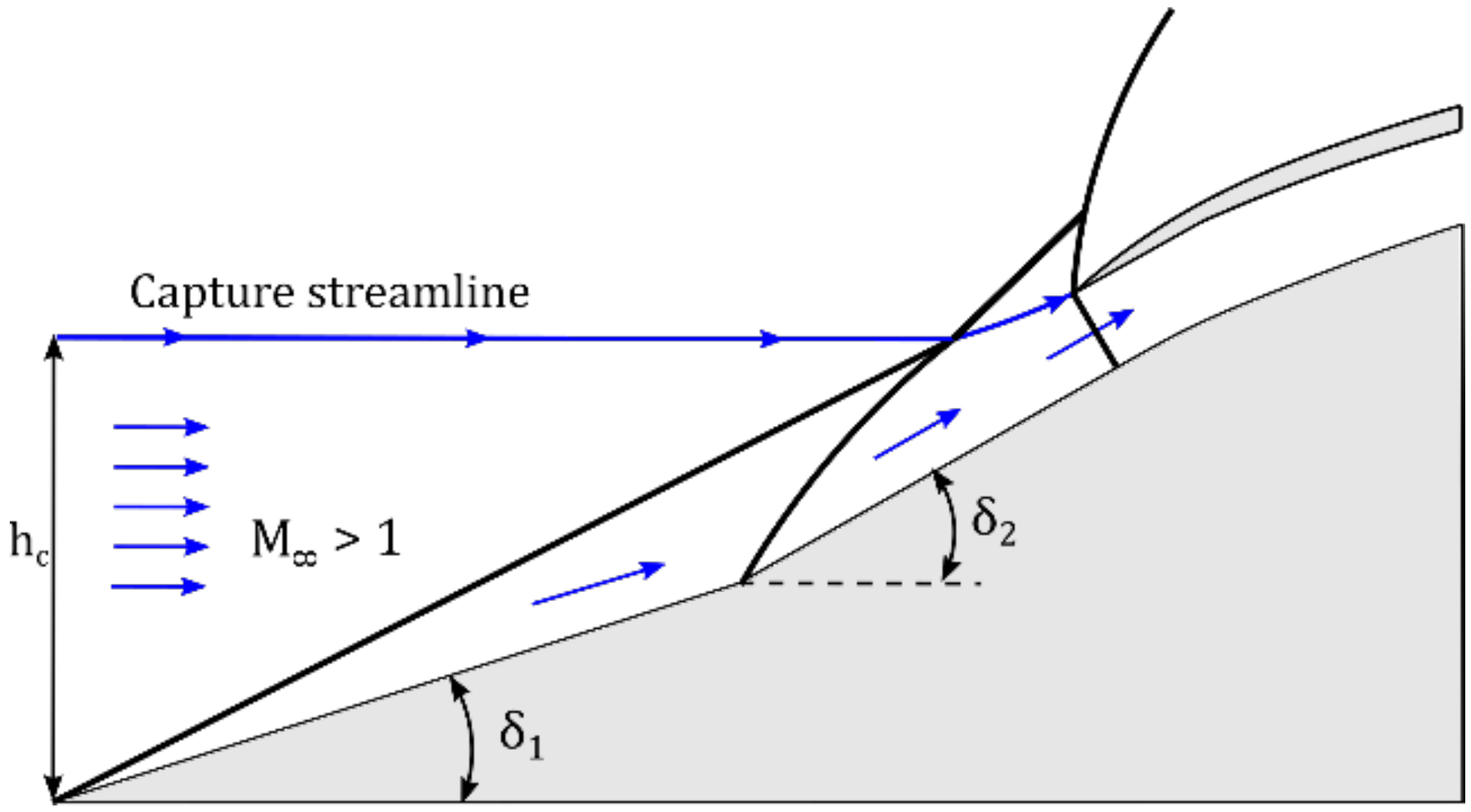

**Fig. 1 Schematic of a biconic intake showing the capture streamline, shock intersection point, and one option of cowl lip location with zero spillage.**

For the problem of maximizing the TPR in the supersonic diffuser segment of the air intake, there is an elegant solution by Oswatitsch [15]. According to the *Oswatitsch criterion*, in a system of (n-1) oblique shocks and one normal shock, the maximum total pressure recovery is obtained when the oblique shocks are of equal strength; that is, when the TPR across each oblique shock is the same. This principle can be easily applied to rectangular intakes, and holds true for conical intakes as well though the calculation in their case is not as straightforward [16]. On the other hand, understandably, estimating the intake drag is a much more complex endeavor, and there are only general guidelines on keeping the intake drag under check. One suggestion is to ensure that the oblique shocks from the intake ramps or cones all meet at the cowl lip, called the *shock-on-lip* condition, though this is possible only at one flight condition. In this case, the capture streamline hits the cowl lip undeflected and the entire capture stream-tube enters the intake duct without any spillage; then the additive drag is zero. In fact, the *shock-on-lip* condition is not strictly necessary; the cowl lip may be displaced aft and radially outwards as shown in Fig. 1, and as long as it is located on the capture streamline, the spillage will still be zero. But, there is now a non-zero additive drag from the curved segment of the capture stream-tube between the shock intersection point and the cowl lip which replaces the cowl drag that would otherwise have accrued from the extension of the cowl lip up to the shock intersection point. There is, thus, a tradeoff between cowl drag and additive drag, but the net drag remains essentially the same [17].

A tradeoff between the mass flow rate and the cowl drag is desirable in cases where a detached bow shock is formed ahead of the cowl lip, which increases the cowl pressure drag. By offsetting the cowl lip marginally in the aft direction such that the ramp or cone shocks converge at a point slightly ahead of the cowl lip, the *shock-on-lip condition* is slightly violated. A small part of the capture stream-tube now spills over the cowl lip, and the free-stream flow no longer directly impinges on the cowl lip, thereby allowing a weaker attached shock to be formed there, as sketched in Fig. 2. Thus, a substantial reduction in cowl pressure drag can be obtained in exchange for a marginal loss in intake mass flow rate and a minor increment in additive drag [18].

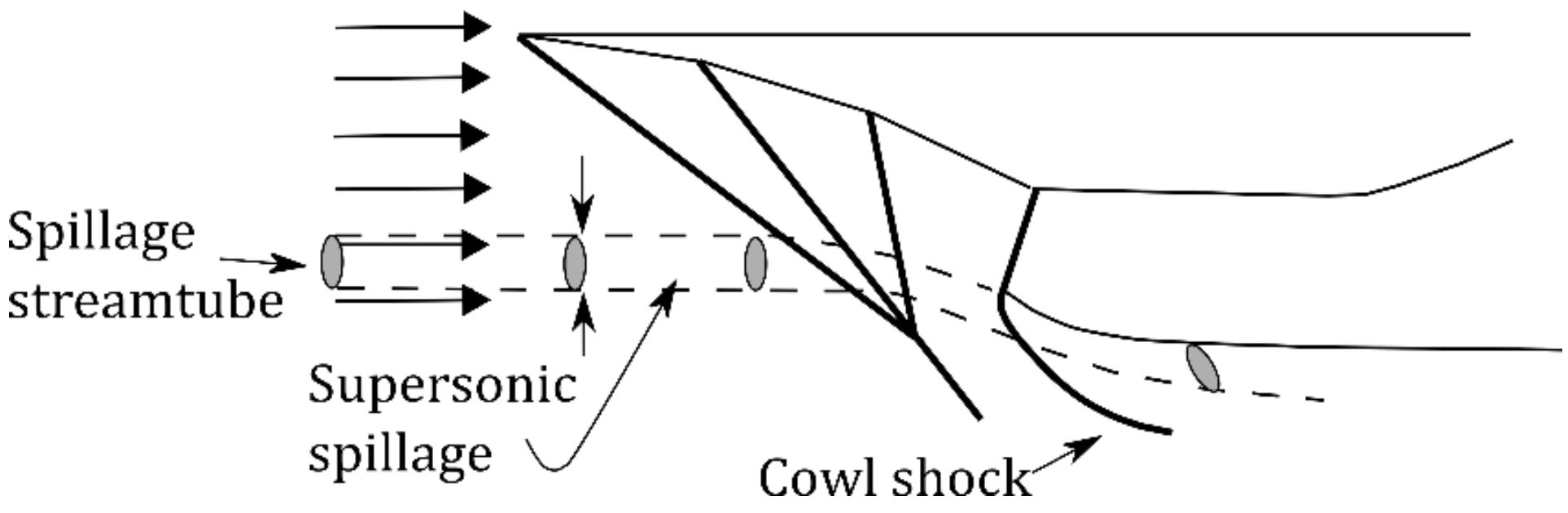

**Fig. 2 Cowl lip aft offset allows for a weaker attached external cowl shock in exchange for a small spillage in mass flow (adapted from Ref. [18]).**

For most supersonic air intakes, the primary tradeoff is between the total pressure recovery (TPR) and the intake drag [19]. The objective of the present work is to devise a systematic methodology to carry out this tradeoff and to quantify the effects thereof for biconic supersonic intake geometries. Starting with the Oswatitsch optimal solution for maximum TPR, our strategy is to lose as little TPR as possible while gaining on drag. To this end, we first introduce the notion of sub-optimal Oswatitsch solutions which are, in a sense, the best combination of ramp or cone angles when the desired TPR is lower than the optimum value. Before beginning the tradeoff between TPR and drag, however, it is desirable to obtain the maximum possible TPR for a biconic intake configuration. In fact, TPR values better than the Oswatitsch-predicted maximum are possible, as demonstrated in Ref. [14], by replacing the terminal normal shock by an alternative shock structure. We, therefore, propose a novel terminal shock arrangement wherein the biconic intake is modified by an additional conical flare, forming what we call a terminal Lambda shock structure including a weaker normal shock. Reynolds-averaged Navier-Stokes (RANS) simulations are carried out to support our hypothesis and reveal that the modified biconic intake does indeed yield a higher TPR than the theoretical Oswatitsch optimum value. Subsequently, we define the optimization problem that forms the basis for the tradeoff between TPR and intake drag, for a fixed value of the intake mass flow rate. The functional requirements, constraints, and design parameters are formulated to satisfy the requirements of Axiomatic Design Theory [20]. The pair of cone angles and the cowl fineness ratio serve as the two design parameters. Two cases are considered: one with a constraint on the intake length, and the other with the intake length unconstrained. Optimal solutions obtained from iterative RANS simulations are plotted in objective-function space to reveal the Pareto front of sub-optimal Oswatitsch solutions. Further, an off-Oswatitsch solution is sought where only the second cone angle deviates from the Oswatitsch-optimal value; this solution is also found to lie on the Pareto front very close to the sub-optimal Oswatitsch solution.

## II. Sub-optimal Oswatitsch Solutions

For a given flight condition, the *Oswatitsch criterion* provides the optimum combination of ramp or cone angles that maximizes the TPR of the supersonic diffuser. The question then arises, how to best choose the ramp or cone angles that give sub-optimal TPR in return for improved intake drag? To address this question, we introduce the concept of sub-optimal Oswatitsch solutions. The procedure for choosing sub-optimal Oswatitsch solutions is illustrated here for the case of a 3-ramp rectangular intake since the Oswatitsch calculations are much simpler in case of planar shocks than for conical shocks. However, the same principle applies to conical shocks as well.

Figure 3 shows contours of constant TPR for a 3-ramp rectangular intake under free stream conditions of Mach number 2.5 and zero angle of attack. In reality, one would have two-dimensional surfaces of constant TPR in the three-dimensional space of incremental ramp angles. For ease of illustration, Fig. 3 shows a two-dimensional slice taken at a constant value of the third ramp angle. The global optimum is the Oswatitsch solution for maximum TPR, marked by a filled square in Fig. 3. If one were to settle for a lower value of TPR in a tradeoff, then it would make sense to select a combination of ramp angles on a TPR contour that is smallest in some sense. Such combinations may be given by the point that is geometrically closest (in a Euclidean sense) to the origin from among all the points on a certain TPR contour. These points form the set of sub-optimal Oswatitsch solutions marked by the dashed line in Fig. 3 and labeled “Oswatitsch Line.” Every point on the “Oswatitsch Line” represents a solution with equal TPR across the first and second ramp shocks, though these TPR values are sub-optimal when compared with the global optimum Oswatitsch TPR. These solutions along the “Oswatitsch Line” are, therefore, called sub-optimal

Oswatitsch solutions. The requirement of "equal TPR" imposes a relationship between the angles $\delta_1$ and $\delta_2$ for a sub-optimal Oswatitsch solution; that is, they are not independent but a pair, $(\delta_1, \delta_2)$. Similarly, in the three-dimensional space of incremental ramp angles, sub-optimal ramp angle triples $(\delta_1, \delta_2, \delta_3)$ may be defined for each two-dimensional TPR surface. In case of triples, if one opts to hold one of the ramp or cone angles fixed for any reason, then the sub-optimal solutions in the space of the remaining ramp or cone angles would be found by taking a two-dimensional slice, in exactly the same manner as indicated in Fig. 3.

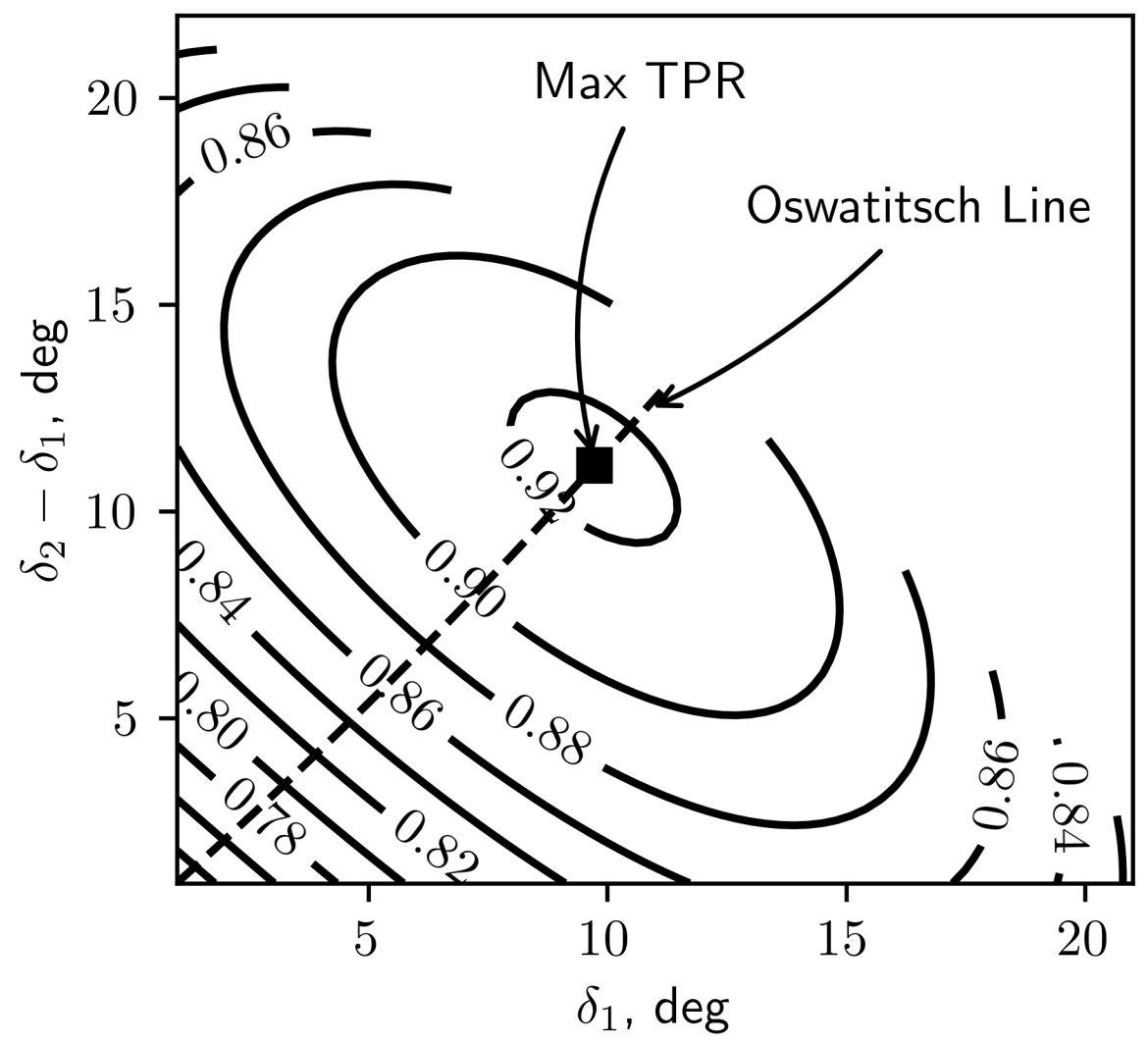

**Fig. 3 Theoretically calculated contours of total pressure ratio in $\delta_1, \delta_2$ space with $\delta_3 = 33$ deg held constant for a 3-ramp rectangular intake at $M_\infty = 2.5$ ($\delta_2 - \delta_1$ is plotted on the Y-axis for better visualization).**

### A. Optimal TPR for biconic intake at Mach 3

Before proceeding further, one needs to determine the optimum cone angles and the Oswatitsch maximum TPR for a biconic intake. The free-stream conditions correspond to Mach 3 at sea level and zero angle of attack. For a first approximation, one may use the Taylor-Maccoll equations [21] and follow a trial-and-error approach for various likely combinations of cone angles. The optimal solution for a biconic external-compression intake so obtained is tabulated in Table 1. The first and second rows are for the two oblique conical shocks, the last row represents the terminal normal shock assumed to be located at the cowl lip. The net TPR is found to be 0.756, which matches the value reported in Refs. [22, 23] precisely. Note the equal TPR across the first and second oblique shocks, as required by the *Oswatitsch criterion*. The effective incidence angle of the flow approaching the second cone is the difference between the average flow turning angle after the first conical shock and the first cone geometric angle. The incremental geometric cone angles are 18.5 and 15.0 deg. The effective second cone angle, however, is larger than its geometric value due to the upstream flow being inclined at a negative effective incidence angle. As a matter of interest, in case of planar shocks in rectangular intakes, the Oswatitsch solution always yields a second ramp angle that is a little larger than the first one. This is completely understandable since the lower Mach number ahead of the second ramp requires a larger ramp angle to create the same static pressure rise and TPR, as required by the *Oswatitsch criterion*. In contrast, the optimal TPR solutions for conical intakes always result in the second geometric cone angle being less than that of the first cone [4, 23]. However, for conical intakes, it is the effective cone angle that determines the flow properties across the second conical shock and, as seen in Table 1, the effective second cone angle is always greater than the first cone geometric angle.

**Table 1 Oswatitsch optimal solution for biconic intake at Mach 3**

| Upstream Mach | Effective incidence angle | Incremental geometric cone angle | Effective cone angle | Downstream Mach | Shock angle | Average flow turning angle | TPR | TPR corrected for curved shock |
|---|---|---|---|---|---|---|---|---|
| 3.0 | 0 | 18.5 deg | 18.5 deg | 2.35 | 28.3 deg | 11.0 deg | 0.953 | 0.953 |
| 2.35 | -7.5 deg | 15.0 deg | 22.5 deg | 1.75 | 36.1 deg | 12.5 deg | 0.961 | 0.953 |
| 1.75 | — | — | — | 0.63 | — | — | 0.833 | 0.833 |

## B. Modified biconic intake with conical flare

It can be observed from Table 1 that a significant loss in total pressure is incurred across the terminal normal shock. One could explore the possibility of obtaining a better net TPR outcome if the terminal normal shock is replaced by an alternative shock structure. This does not violate the *Oswatitsch criterion* — the Oswatitsch solution merely assumes a single terminal normal shock. With an alternative terminal shock structure, the upstream ramp or cone oblique shocks still obey the Oswatitsch condition, but the value of the net TPR behind the revised terminal shock may be different, preferably higher. In case of rectangular intakes, such a solution was successfully devised by us in Ref. [14] where an internal wedge angle was created at the cowl. This replaced the terminal normal shock by a strong oblique shock, which afforded a better TPR, with subsonic flow downstream. Unfortunately, the strong form of the oblique shock is absent in conical flows. Instead, we propose here a novel solution where the terminal normal shock is replaced by what we call a Lambda shock (not to be confused with the lambda shock in shock-wave boundary-layer interaction [24]). To this end, we introduce a conical flare downstream of the second cone, as indicated in Fig. 4, where the oblique shock from the conical flare hits the inside of the cowl, and is followed by a much weaker terminal normal shock. The combination of the conical flare shock and the revised normal shock (which resembles the Greek letter $\lambda$) may be expected to yield a higher TPR while leaving the external flow structure of the two conical shocks coalescing into a single external cowl shock unchanged. Hence, neither the cowl pressure drag nor the intake mass flow is altered by this modification. It must be emphasized that the modified biconic intake with the conical flare is different from a triple cone intake where the conical shock from the third cone would be expected to converge at the common shock intersection point and obey the Oswatitsch condition itself. That would yield a higher TPR but at the cost of increased drag since a stronger coalesced shock would be formed at the cowl resulting in higher pressure on the cowl external surface. In contrast, the introduction of the conical flare in Fig. 4 only alters the internal flow through the intake duct leaving the external flow over the cowl unaffected. The performance of the modified biconic configuration with the terminal Lambda shock is verified in the next section by numerical simulation.

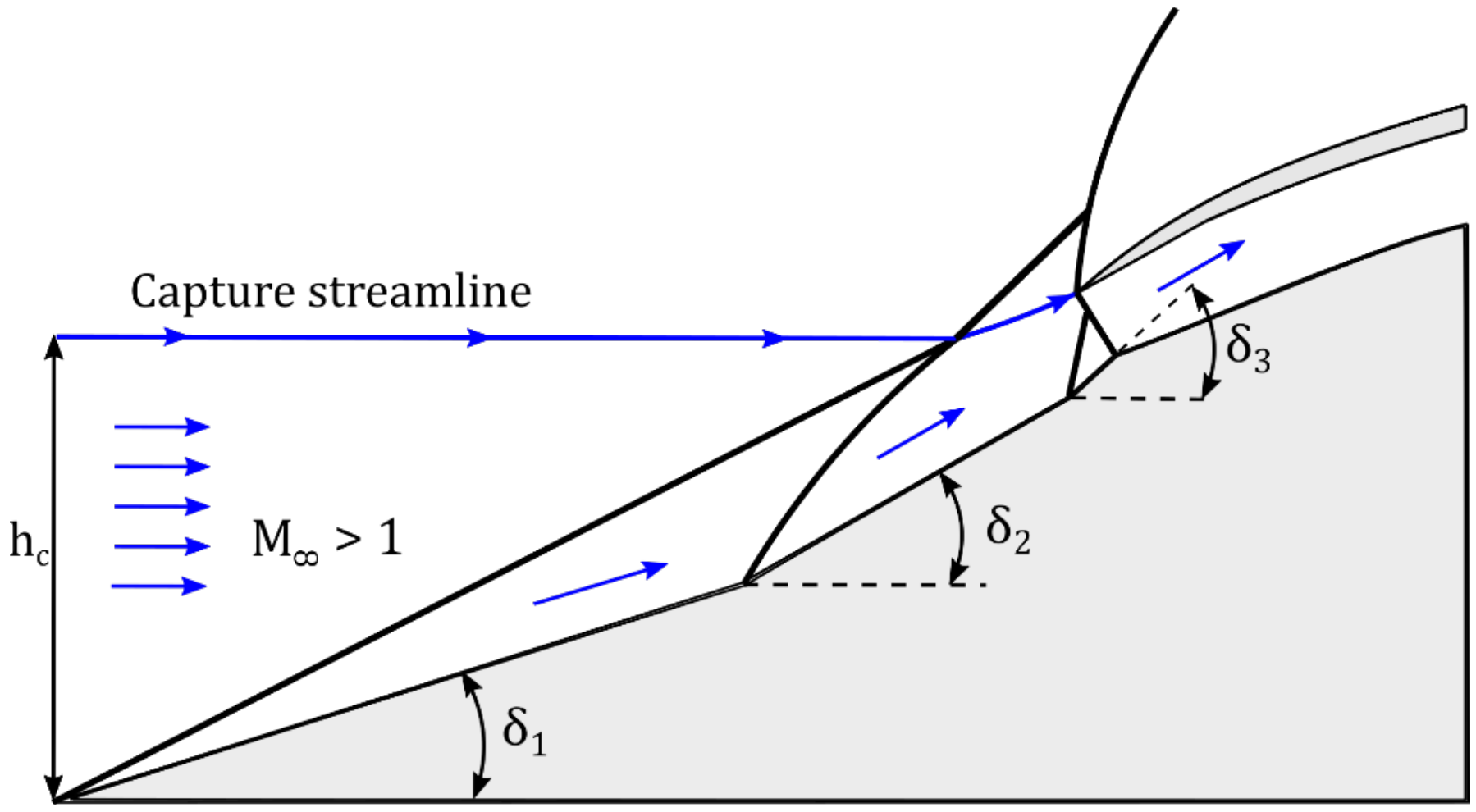

**Fig. 4  Schematic of the modified biconic intake with conical flare showing the proposed Lambda terminal shock structure.**

## III. Numerical Simulation Procedure

The modified biconic intake configuration in Fig. 4 with first cone angle 18.5 deg as per Table 1 is numerically simulated to evaluate its total pressure recovery (TPR). The free-stream flow conditions correspond to Mach 3.0 at sea level and zero angle of attack. The capture stream-tube is sized for a mass flow rate of 4.5 kg/s with an error margin of 2%; the cowl lip is positioned accordingly such that there is no flow spillage. The second cone angle is determined iteratively to satisfy the *Oswatitsch criterion*, that is, equal TPR across the first and second conical shocks. The conical flare angle is then determined such that the shock from the flare impacts just inside the cowl. The cowl external surface has an ogive shape similar to that in Ref. [25]. The computational domain definition and boundary conditions imposed are as indicated in Fig. 5. Supersonic inlet conditions are used at the free-stream, supersonic outlet conditions are applied at the far-field, and the intake exit is modeled as a subsonic outlet where the static pressure is specified. All walls, internal and external, are specified to be no-slip, adiabatic surfaces. The simulation uses a mixed-element mesh for the 2-D axisymmetric domain in Fig. 5, with a higher density of grid points in regions of expected sharp gradients, that is, near the conical spikes and around the cowl. The boundary layer region is resolved using a wall-normal grid of 44 layers with the mesh containing approximately 200,000 cells in total.

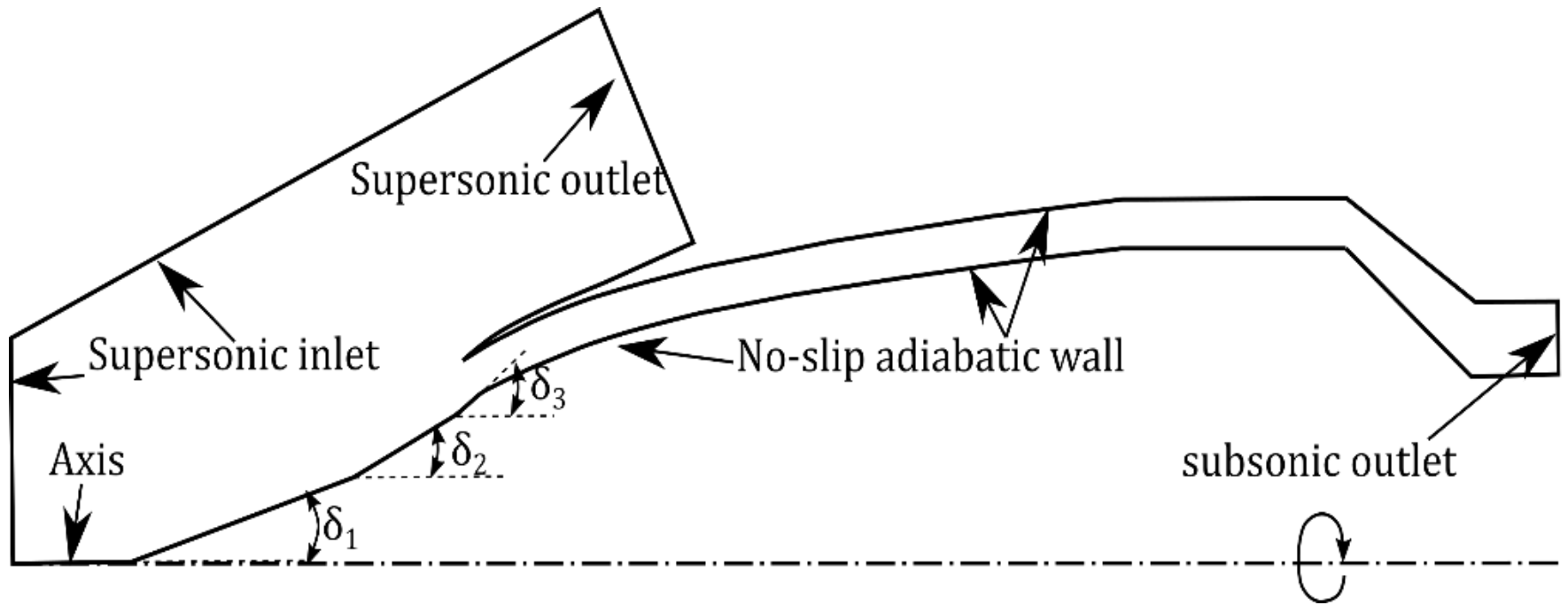

**Fig. 5 Definition of the computational domain for modified biconic intake with boundary conditions indicated.**

The SU2 code [26], a suite of open-source software tools for the numerical solution of Computational Fluid Dynamics (CFD) problems, is used for this study. The governing equations are the axisymmetric compressible Reynolds-averaged Navier-Stokes (RANS) equations in cylindrical polar coordinates, for conservation of mass, momentum, and energy, as given, for example, in Ref. [27]. A standard ideal gas is assumed as the working fluid, with specific heat ratio $\gamma = 1.4$, and specific gas constant $R = 287.058$ J/kg-K. The system of governing equations is closed using the caloric equation of state to determine pressure, and the thermal equation of state to determine static temperature. Sutherland's law is used to determine the local value of dynamic viscosity using the local temperature [28]. Conductivity is dependent on the specific heat at constant pressure $C_p$, the Prandtl number ($Pr = 0.72$), and the turbulent Prandtl number ($Pr_t = 0.9$). The solver uses the AUSM scheme [29] for inviscid flux reconstruction, along with the higher-order MUSCL method [30] and the Venkatakrishnan limiter [31]. The Euler implicit method is used for time integration. Menter's Shear Stress Transport (SST) turbulence model [32] is used, and a Y+ of less than 1.0 is maintained while generating the meshes since the SST model does not use a wall function. Simulations are taken to have converged when the root-mean-square (RMS) residuals have fallen by at least four orders. Grid convergence study is presented in Appendix A.

The intake total pressure recovery (TPR) is obtained as the ratio of the mass-averaged total pressure at the intake exit plane to the free-stream total pressure. The mass flow rate entering the intake duct is evaluated by integrating at a plane slightly behind the terminal shock at the duct entrance. The same plane may be used to estimate the TPR of the supersonic diffuser segment alone. Additive drag is calculated by integrating the pressure over the surface of the capture stream-tube, while both pressure and skin-friction drag are evaluated for the ogive-shaped external cowl surface. As a further check, additive drag is also computed by integrating the pressure and skin friction over the cone surface, and accounting for the pressure and momentum terms at the entrance to the intake duct at the cowl-lip station and at the free-stream [12]. Both methods are verified to yield the same value of additive drag within a small margin of numerical error. The drag coefficient is computed using the free stream flow properties with the area of circular base of the intake as the reference area.

**Table 2 Computational Oswatitsch optimal solution for modified and original biconic intake at Mach 3**

| Property | Modified biconic intake | Original biconic intake |
|---|---|---|
| Cone and flare angles, deg | 18.5, 31.0, 41.0 | 18.5, 31.0 (no flare) |
| Additive drag coefficient | 0.024 | 0.025 |
| Cowl drag coefficient | 0.210 | 0.210 |
| Net intake drag coefficient | 0.234 | 0.235 |
| Critical back-pressure ratio | 22.1 | 18.7 |
| Exit Mach Number | 0.34 | 0.39 |
| TPR at exit | 0.645 | 0.562 |

The Oswatitsch optimal solution for the modified biconic intake is found to occur for cone angles of 18.5 and 31.0 deg, and flare angle of 41.0 deg. That is, the incremental second cone angle is 12.5 deg, slightly different from the approximate value of 15.0 deg estimated in Table 1. This value is independent of the presence or absence of the conical flare modification. That is, the Oswatitsch optimal cone angles for the original biconic intake are the same as well. The modified biconic intake shows a supersonic diffuser TPR of 0.811 under viscous flow conditions, well above the theoretical inviscid Oswatitsch value of 0.756, confirming the value of replacing the conventional single terminal normal shock with the novel Lambda terminal shock structure. On the other hand, the expected viscous-flow TPR for the supersonic diffuser with the original biconic intake is approximately 0.707, understandably somewhat lower than the theoretical best value of 0.756.

Other properties of both the modified and original optimal intake configurations are listed in Table 2. In each case, the back-pressure ratio is adjusted until the terminal normal shock is located at the cowl lip; that is, the intakes are both in critical operation. When considering the intake as a whole up to the subsonic exit, the modified biconic geometry offers a much better optimal TPR of 0.645 as against the value of 0.562 for the original biconic intake — that is, a nearly 15 per cent increase in TPR. Also, both the biconic intakes in Table 2 show almost identical values of intake drag. Thus, it is apparent that the modified biconic intake with an additional conical flare is a much better starting point for a tradeoff between TPR and intake drag.

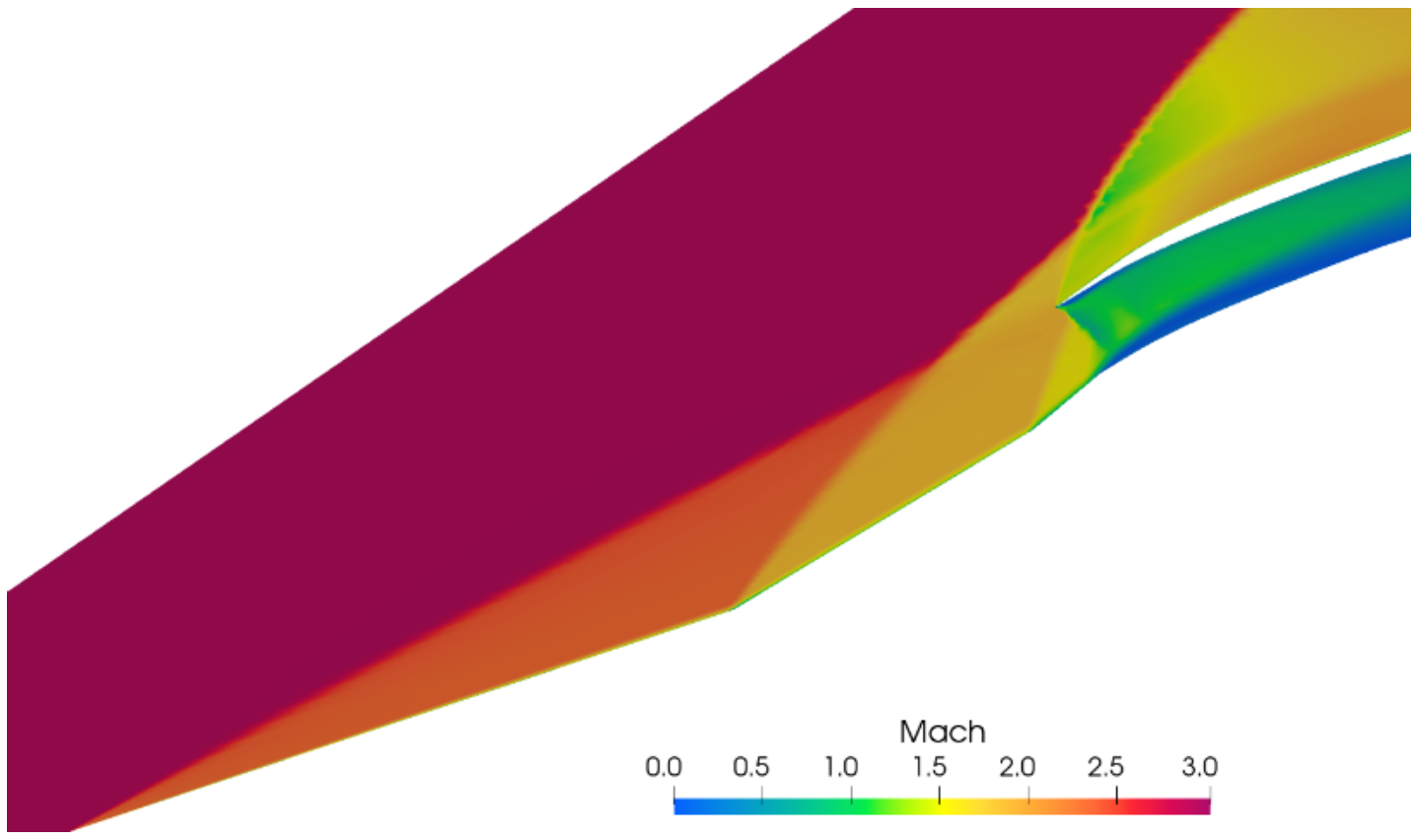

**Fig. 6 Mach number contours for the modified biconic intake design at free-stream Mach number 3.0, zero angle of attack, and critical back-pressure ratio 22.1.**

Mach contours for the flow past the modified biconic intake are shown in Fig. 6. The two conical shocks are seen to intersect at a point. The cowl lip is offset from the shock intersection point, but lies on the capture streamline as sketched in Fig. 4; there is, thus, no flow spillage at the cowl lip. The external flow over the cowl has a lower upstream Mach Number (as compared to the free-stream Mach number), creating a weak attached shock at the cowl lip which then merges with the coalesced conical shock. The interaction between these two shocks traveling in the same direction results in a contact discontinuity that is clearly visible [33]. The conical shock due to the flare angle, forming the first leg of the terminal Lambda shock arrangement, impacts the terminal normal shock which is at its critical position. In case of super-critical operation of the intake (back-pressure ratio lower than the 22.1 value for Fig. 6), the flare conical shock would reflect from the inside of the cowl surface with the terminal normal shock displaced slightly downstream. There is a region of separated flow originating near the foot of the terminal normal shock and extending downstream into the intake duct. This is entirely in keeping with observations in the literature and is a characteristic of intake buzz as per the Dailey criterion [34].

As a matter of abundant caution, similar simulations for the Oswatitsch optimal solution are carried out for first cone angles of 17.5 and 19.5 deg, respectively, to compare the best TPR obtained in those cases with the values reported in Table 2. The optimal TPR in either case is found to be of the order of 0.01 lower than that for the 18.5 deg first cone case, confirming the 18.5 deg solution to be the optimal one.

## IV. Optimization with Intake Length Unconstrained

The geometrical features of biconic intakes that are of interest in the formulation of the optimization problem are first noted. With reference to Fig. 7, besides the cone half-angles $\delta_1$ and $\delta_2$, the other parameter of interest is the Fineness Ratio (FR), defined as the ratio of the imaginary fore-body length $L$ and the intake base diameter $D$. The imaginary fore-body is created by extending the ogival shape of the cowl upstream, as sketched in Fig. 7 by the dashed curve. The Fineness Ratio is a key parameter that determines the wave drag of the fore-body in supersonic flow, hence it directly correlates with cowl drag for the biconic intake [35]. The actual length of the intake is the length of the conical spikes plus the length of the cowl section, which is marked in Fig. 7.

There are four factors of interest in the optimization problem as applied to the design of biconic supersonic intakes: 1. The total pressure recovery (TPR) between the free-stream flow and the intake exit station, 2. The net intake drag, comprising of additive and cowl drag, 3. The intake mass flow rate, and 4. The intake length, consisting of the cone length and the cowl length. At the same time, there are four parameters at hand: The two cone half-angles $\delta_1$ and $\delta_2$, the Fineness Ratio (FR), and the location of the cowl lip. The cone angles $\delta_1$ and $\delta_2$ obviously determine the TPR, as well as influence the cowl drag via the static pressure behind the external shock structure acting on the outer cowl surface. In addition, they decide the cone part of the intake length. The Fineness Ratio has a slight influence on the TPR (as explained later), but directly affects the cowl pressure drag, and also the cowl length. The location of the cowl lip regulates the intake mass flow rate and influences the cowl length as well. Thus, there are significant cross-relations between the four available parameters and the four factors of interest, making the formulation of the optimization problem a non-trivial one.

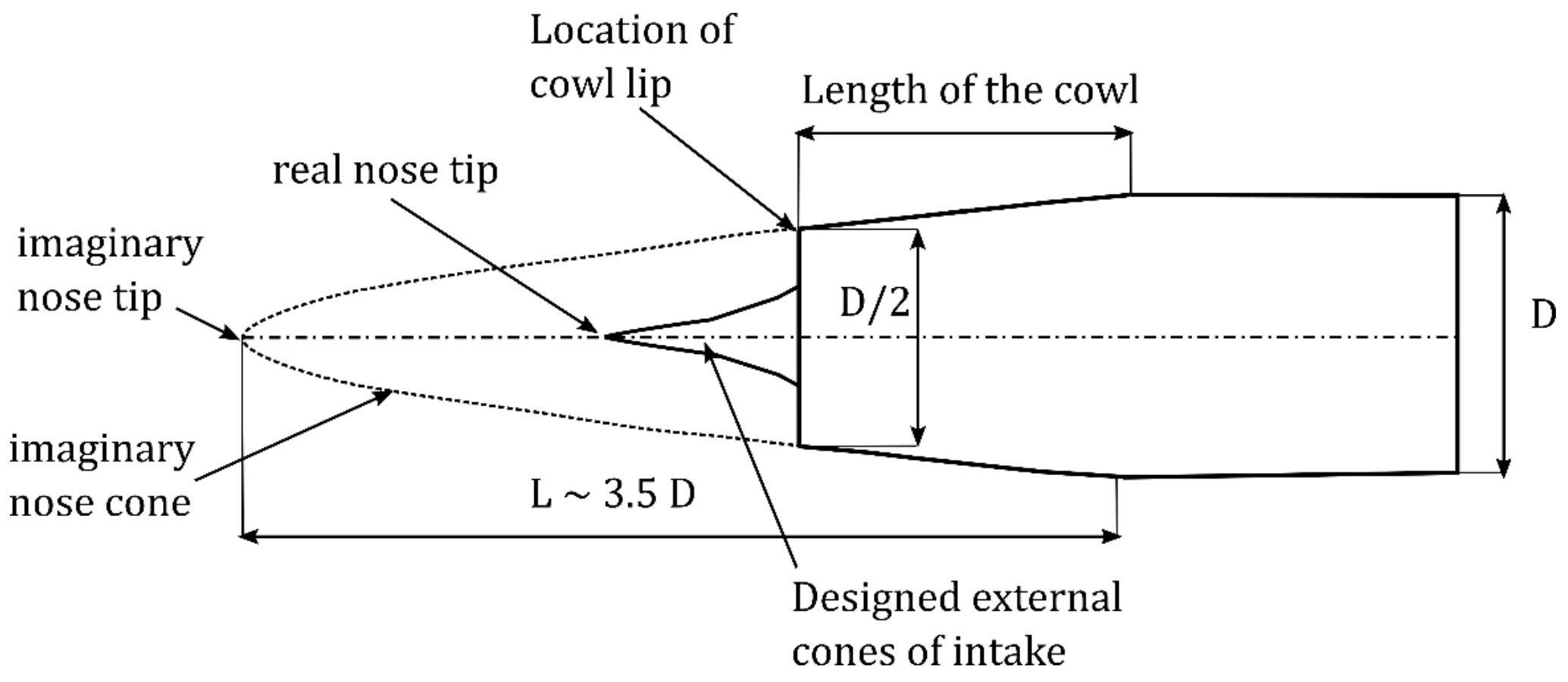

**Fig. 7 Geometrical features of biconic intakes including the imaginary fore-body that is used to define the Fineness Ratio, *FR=L/D*.**

### A. Formulation

In order to ensure that the optimization problem is well formulated, we shall follow the guidelines of Axiomatic Design Theory (ADT) [20]. ADT is concerned with the objective functions (OF), design parameters (DP), and the mapping between the DP and OF, from physical space to functional space, represented by the Design Matrix (DM). There are a few guidelines from ADT which may be paraphrased as follows: 1. Minimize the number of OF as far as possible, with the other requirements being stated as constraints, 2. Ensure that the OF are mutually independent, 3. Choose as many DP as there are OF in the formulation, 4. Ideally, choose the DP such that the design is uncoupled, that is, each DP affects only one unique OF — that would make the Design Matrix (DM) diagonal. However, if that is not possible, at least ensure that the DM is triangular, which allows for the system to be solved sequentially.

In our case, there are two primary objective functions: 1. Maximize TPR, and 2. Minimize net intake drag. The mass flow rate requirement can be imposed as a constraint. For the present section, the intake length is unconstrained, hence that is not a factor of interest. Of the four parameters at hand, the location of the cowl lip may be devoted to meeting the mass flow rate constraint, which leaves three parameters — the two cone angles and the FR — to meet only two objective functions. In order to meet the guidelines of ADT, the three parameters will need to be reduced to two; however, clearly none of them can be neglected. This is where the notion of sub-optimal Oswatitsch solutions introduced in Sec. II comes handy. In that case, the angles $\delta_1$ and $\delta_2$ are not independent, but are related by the "equal TPR" requirement, which means that the pair $(\delta_1, \delta_2)$ may be considered to be a single entity. The relationship between the DP and OF may then be symbolically written as:

$$\begin{bmatrix} TPR \\ Cd_i \end{bmatrix} = \begin{bmatrix} \checkmark & \epsilon \\ \checkmark & \checkmark \end{bmatrix} \begin{bmatrix} (\delta_1, \delta_2) \\ FR \end{bmatrix} \tag{1}$$

where $Cd_i$ is the net intake drag, the checkmarks indicate a dependence, and the '$\epsilon$' indicates that TPR is a relatively weak function of the Fineness Ratio (FR). The formulation in Eq. (1) may be seen to meet all the guidelines of ADT as stated above, with the Design Matrix (DM) being approximately lower triangular.

### B. Solution

Ideally, the solution of the optimization problem can be automated by coupling an optimizer with a RANS solver. Starting off with an initial selection of test points in Design Parameter (DP) space, additional infill points may be introduced at every iteration of the optimizer until a converged Pareto front is obtained. The Pareto front represents the set of non-dominated solutions — solutions for which it is impossible to improve any OF without simultaneously degrading one or more of the other OF [36]. Thus, the Pareto front gives the set of solutions between which a tradeoff may be conducted. Such an exercise was carried out by us in Ref. [14] for a multi-ramp rectangular intake. However, automation with a RANS solver in the loop is difficult in case of biconic intakes. As seen previously, the actual value of the second cone angle (12.5 deg in Table 2) differed from the theoretically calculated one (15.0 deg in Table 1). Further, in case of the modified biconic intake, the flare angle will have to be manually adjusted so that the flare shock impacts the cowl just inside of the cowl lip. Since the second and flare shocks are curved and theoretical predictions are unreliable, the solution must be arrived at by numerical experimentation. Also, to compare the best TPR values between the various solutions in a fair manner, the intake in each case must be in critical operation. That is, iterations with increasing values of back-pressure need to be done to ensure that the final Lambda shock has the normal shock in a critical state. For these reasons, the solution process is not automated; instead the optimal solutions are obtained manually as explained next.

The solution process takes advantage of the nearly triangular form of the Design Matrix (DM) in Eq. (1) whereby the TPR may first be determined by setting the cone angles $(\delta_1, \delta_2)$. Subsequently, $Cd_i$ can be optimized by adjusting the FR — changes to the FR have only a minor impact on the TPR according to the formulation in Eq. (1).

Broadly, the solution process is as follows: For a given $\delta_1$, iteratively determine the $\delta_2$ with equal TPR — this is the sub-optimal cone angle combination. Then create the desired conical flare angle to place the first part of the terminal Lambda shock in place. The next step is to adjust the FR to optimize the cowl drag. In general, increasing the FR will improve the cowl drag, but the choice of FR is limited by the cowl angle at the lip, which is constrained by the mass flow rate requirement. Finally, increase the back-pressure until the terminal normal shock is located at the cowl lip, setting the second part of the terminal Lambda shock in place. Minor refinements may still be necessary to account for unexpected viscous effects.

**Table 3 Sub-optimal Oswatitsch solutions for modified biconic intake at Mach 3**

| Property | Case 1 | Case 2 |
|---|---|---|
| Cone and flare angles, deg | 12.0, 18.0, 26.0 | 15.0, 23.5, 33.0 |
| Additive drag coefficient | 0.005 | 0.010 |
| Cowl drag coefficient | 0.084 | 0.103 |
| Net intake drag coefficient | 0.089 | 0.113 |
| Critical back-pressure ratio | 19.5 | 19.6 |
| Exit Mach Number | 0.42 | 0.42 |
| TPR at exit | 0.598 | 0.606 |
| Fineness Ratio | 3.0 | 2.5 |
| Intake length, mm | Baseline + 100 | Baseline + 50 |

Sub-optimal Oswatitsch solutions obtained in this manner for two cases, with first cone angle of 12.0 deg (Case 1) and 15.0 deg (Case 2), respectively, are presented in Table 3, with mass flow rate constrained to 5.0 kg/s with an error margin of 2%. The tradeoff between TPR and the intake drag coefficient $Cd_i$ between Cases 1 and 2 in Table 3, and relative to the Oswatitsch optimal design in Table 2, is apparent. Compared to the optimal TPR of 0.645 for the 18.5 deg design in Table 2, the TPR in Cases 1 and 2 are only around 0.6. However, there is a whopping decrease in the intake drag coefficient from 0.234 for the Oswatitsch optimal design of Table 2 to 0.113 for Case 2, and further down to 0.089 for Case 1. The relatively large decrease in $Cd_i$ as the cone angles are reduced in exchange for a smaller percentage loss in TPR is typical when the Pareto front has a hyperbolic shape; similar results have been noted for rectangular intakes in Ref. [14]. However, lowering the cone angles as well as increasing the FR have together resulted in a significant increase in the intake length for the cases in Table 3. Nevertheless, the results in Table 3 reinforce the conclusion drawn in Ref. [14] that the best choice of cone angles in case of a multi-objective optimization of supersonic intakes is preferably somewhat smaller than that predicted by the *Oswatitsch criterion*. As shown in Sec. II, these cone angles may be selected from among the sub-optimal Oswatitsch solutions as described in Fig. 3.

## V. Solutions with Intake Length Constraint

In practice, the intake length is likely to be limited due to various other considerations in vehicle design. Hence, it is of interest to investigate possible tradeoffs between TPR and intake drag with a constraint placed on the intake length. In this section, solutions are sought with the intake length constrained to Baseline + 50 mm. All other parameters are unchanged from the analysis in the previous section. Examining Eq. (1), once the sub-optimal pair $(\delta_1, \delta_2)$ is chosen, it determines the TPR as well as the cone part of the intake length. That leaves the FR as the only remaining parameter to fix the cowl part of the intake length such that the intake length constraint is met, which means that there is no free parameter to regulate $Cd_i$. In other words, the OF of minimizing intake drag is no longer tenable; the value of $Cd_i$ is naturally obtained once the other parameters are set. Thus, strictly, no tradeoff between TPR and $Cd_i$ may be engineered in this scenario, but it is worth investigating a few cases nevertheless. Table 4 presents the length-constrained solutions for three cases — Case 4 is the same as Case 2 in Table 3 and is the point of comparison.

Case 3 uses the same sub-optimal cone angles of (12.0, 18.0, 26.0) deg as Case 1 in Table 3, except that the length has been shortened by 50 mm in Case 3. Since the cone length does not change between Cases 1 and 3, the length reduction must take place entirely in the cowl; hence, the FR is reduced from 3.0 to 2.3. Consequently, $Cd_i$ increases whereas the TPR is unchanged, correctly reflecting the relation between FR and $Cd_i$ in Eq. (1) and the negligible impact of FR on TPR. Comparing Cases 3 and 4 in Table 4, Case 3 has a lower TPR and a higher $Cd_i$, making it unattractive on both counts. There is, therefore, no tradeoff between the two objective functions in Cases 3 and 4 — Case 4 is superior on each count.

Case 5 has the same optimal cone angles of (18.5, 31.0, 41.0) deg as in Table 2 but the intake length has now been increased from the baseline value in Table 2 by 50 mm. Again, the entire increase occurs in the cowl length with the FR increasing from 1.9 to 2.6. As a result, one would expect the cowl drag in Case 5 to decrease, which is exactly what is observed in Table 4. However, there is an additional complication in this case. Due to the higher FR, the cowl angle at the lip has decreased; on the other hand, due to the large cone angles, the flow deflection angle behind the second conical shock is high. Together, these two effects cause the flow incidence angle at the cowl lip to be high, which creates an internal oblique shock where the terminal normal shock should have formed. If this flow incidence angle exceeds the limit for attached oblique shocks, then a detached shock is generated at the cowl lip. Consequently, there is a loss in TPR and a sharp increase in additive drag. This is an instance where a change in FR actually has an effect on the TPR, and the reason why a small '$\epsilon$' was introduced in Eq. (1). Comparing Cases 4 and 5 in Table 4, Case 5 has a higher net $Cd_i$ and a worse TPR, making it inferior to Case 4 on both parameters.

**Table 4 Sub-optimal Oswatitsch solutions for modified biconic intake at Mach 3 with length constraint**

| Property | Case 3 *(sub-optimal cone angles)* | Case 4 *(sub-optimal cone angles)* | Case 5 *(Oswatitsch-optimal cone angles)* |
|---|---|---|---|
| Cone and flare angles, deg | 12.0, 18.0, 26.0 | 15.0, 23.5, 33.0 | 18.5, 31.0, 41.0 |
| Additive drag coefficient | 0.009 | 0.010 | 0.033 |
| Cowl drag coefficient | 0.127 | 0.103 | 0.091 |
| Net intake drag coefficient | 0.136 | 0.113 | 0.124 |
| Critical back-pressure ratio | 19.5 | 19.6 | 18.2 |
| Exit Mach Number | 0.42 | 0.42 | 0.41 |
| TPR at exit | 0.598 | 0.606 | 0.576 |
| Fineness Ratio | 2.3 | 2.5 | 2.6 |
| Intake length, mm | Baseline + 50 (Shortened from Case 1) | Baseline + 50 (Same as Case 2) | Baseline + 50 (Lengthened from TPR-optimal baseline) |

Thus, the sub-optimal design in Case 4 is the one that offers the best tradeoff between TPR and $Cd_i$ with the length constraint of Baseline + 50 mm. More generally, in may be stated that, for any constrained value of intake length, the corresponding sub-optimal Oswatitsch solution, as in Table 3, would be the best option for a tradeoff between TPR and $Cd_i$.

### A. Off-Oswatitsch solution

There is an alternative solution possible for the length-constrained problem by decoupling the pair $(\delta_1, \delta_2)$. Instead of losing TPR by reducing both the cone angles according to the sub-optimal Oswatitsch solutions, one can hold the first cone angle at its Oswatitsch optimum value of 18.5 deg and decrease only the second cone angle. In this case, the two parameters $\delta_2$ and FR are jointly used to meet the length constraint as follows:

$$[\text{Length of cone part} : f_1(\delta_2)] + [\text{Length of cowl part} : f_2(FR)] = [\text{Intake length} : \text{Baseline} + 50 \text{ mm}] \tag{2}$$

The same two parameters also influence $Cd_i$, as seen in Eq. (1). The problem, then, is to minimize $Cd_i$ subject to the condition in Eq. (2). Formulated in this manner, the OF to maximize TPR is abandoned; the TPR is naturally obtained once the $Cd_i$ minimization problem yields a value of $\delta_2$ (and the appropriate value of the flare angle with it). Unfortunately, this formulation goes against the tenets of ADT since the same two parameters are used to satisfy a constraint as well as optimize an OF. Nevertheless, a series of numerical iterations can be conducted to reach an approximate optimal solution.

**Table 5 Off-Oswatitsch solution for modified biconic intake at Mach 3 with length constraint**

| Property | Case 5 *(Oswatitsch-optimal cone angles)* | Case 6 *(off-Oswatitsch second cone angle)* |
|---|---|---|
| Cone and flare angles, deg | 18.5, 31.0, 41.0 | 18.5, 25.0, 35.0 |
| Additive drag coefficient | 0.033 | 0.005 |
| Cowl drag coefficient | 0.091 | 0.099 |
| Net intake drag coefficient | 0.124 | 0.104 |
| Critical back-pressure ratio | 18.2 | 19.7 |
| Exit Mach Number | 0.41 | 0.42 |
| TPR at exit | 0.576 | 0.604 |
| Fineness Ratio | 2.6 | 2.6 |
| Intake length, mm | Baseline + 50 | Baseline + 50 |

Case 6 in Table 5 lists the intake and flow properties for a modified biconic design with first cone angle as 18.5 deg and an off-Oswatitsch second cone angle of 25.0 deg, which yields the minimum $Cd_i$ of 0.104. This is compared with Case 5 in Table 4, also having a first cone angle of 18.5 deg, but with the Oswatitsch-optimal second cone angle of 31.0 deg. The incremental conical flare angle of 10 deg is the same in both cases. The lower second cone angle in Case 6 reduces the flow incidence angle approaching the cowl lip so that the internal cowl lip shock is once again a normal shock. That is, the terminal Lambda shock structure is recovered and the external shock at the cowl lip is now attached. Consequently, the increased additive drag in Case 5 is not seen in Case 6, and the TPR has also improved. Case 6 out-performs the Case 5 design on $Cd_i$ as well as TPR. At this stage, it would appear that the off-Oswatitsch solution is a strong alternative to the sub-optimal Oswatitsch solutions considered in Table 4.

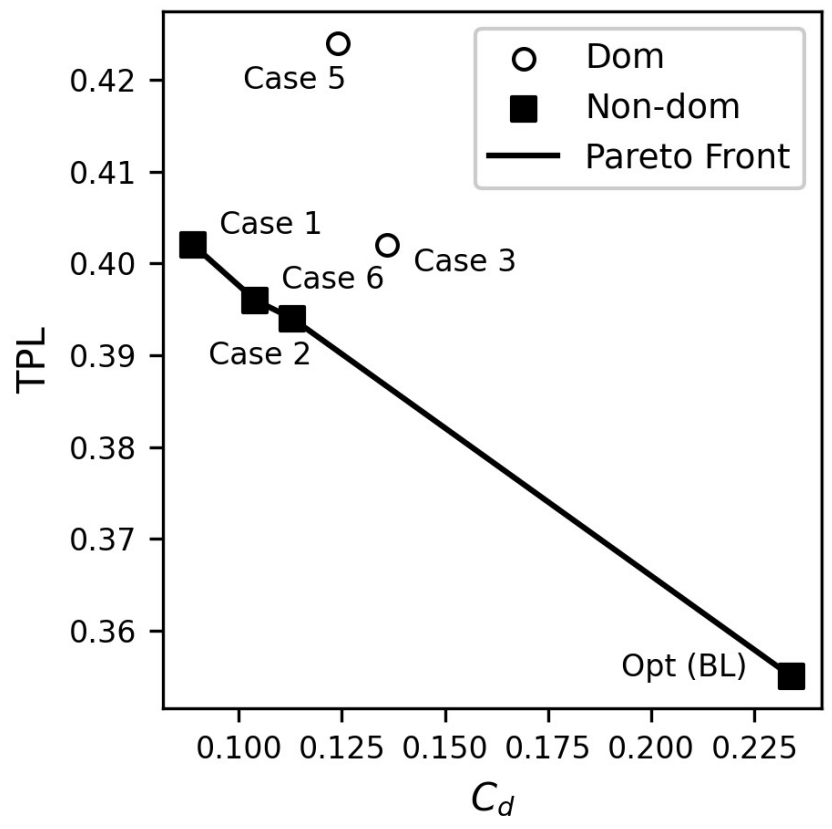

**Fig. 8 Solutions in Cases 1 through 6 plotted in objective-function space with the Pareto front.**

## B. Pareto front

To place the solutions in Cases 1 through 6 in perspective, it is useful to mark them on a plot of Total Pressure Loss (TPL = 1 - TPR) vs. $Cd_i$, as shown in Fig. 8 (Case 4 is the same as Case 2). The TPR-optimal Oswatitsch solution has the baseline intake length. Case 1 alone has intake length of Baseline + 100 mm, all the others are of Baseline + 50 mm length. Only Case 6 has off-Oswatitsch cone angles, all the others are either Oswatitsch-optimal or sub-optimal cones. Four of the solutions are non-dominated (filled black squares), implying that there is no other solution that improves on any of them simultaneously in both respects, TPR and $Cd_i$. Hence, they form the Pareto front. Along the Pareto front, one can trade off TPR for $Cd_i$, albeit with intake length increasing from Baseline (“Opt (BL)”) to Baseline + 100 mm (Case 1). If a fixed value of intake length is desired, then there is only one sub-optimal

Oswatitsch solution along the Pareto front, with a fixed tradeoff between TPR and $Cd_i$, such as Case 2 for intake length of Baseline + 50 mm. However, other solutions, such as the off-Oswatitsch Case 6 offer a slightly different tradeoff between TPR and $Cd_i$ for the same intake length. In numerical terms, Case 2 and Case 6 are quite close, suggesting that no significant advantage may be gained by exploring alternative solution methodologies to the sub-optimal Oswatitsch solutions. In that case, the sub-optimal Oswatitsch solutions may be considered to be the accepted procedure to trade off between TPR and $Cd_i$ for supersonic intakes.

## VI. Conclusion

A systematic approach to carrying out the tradeoff between the competing objectives of maximizing TPR and minimizing $Cd_i$ for supersonic intakes has been presented in this work. In the case of biconic intakes, the corresponding design parameters are selected to be the pair of cone angles $(\delta_1, \delta_2)$ and the cowl fineness ratio (FR). The cone angle pairs $(\delta_1, \delta_2)$ are chosen such that their respective conical shocks satisfy the "equal TPR" condition — such solutions have been labeled as sub-optimal Oswatitsch solutions. The set of sub-optimal Oswatitsch solutions is shown to form the Pareto front in the objective-function space. Thus, the tradeoff between TPR and $Cd_i$ may be conducted by switching between the sub-optimal Oswatitsch solutions along the Pareto front. However, the length of the intake is variable in this process. With the intake length constrained, it is seen that no tradeoff is really possible between TPR and $Cd_i$ — there is one sub-optimal Oswatitsch solution on the Pareto front with the desired intake length. An alternative method using what is called an off-Oswatitsch solution is tested for the length-constrained case, but it yields a solution very nearly the same as the sub-optimal Oswatitsch solution with the same value of the intake length.

## Appendix A: Grid Convergence

A grid convergence study is undertaken to determine the optimal grid resolution for the computational analysis. The baseline modified biconic intake case geometry with conical flare, having angles (18.5,31.0,41.0) deg, is selected for this analysis. Three different grid levels are considered, namely, Coarse (L0), Medium (L1), and Fine (L2), with details provided in Table 6. To ensure appropriate resolution of the boundary layer, near-wall viscous layers are incorporated into the meshes. The Mach number along a streamline passing through the center of the capture stream-tube, originating upstream of the first ramp and passing wholly into the intake duct, is plotted in Fig. 9 to compare the results across different grid levels. The predictions of L1 and L2 meshes show good agreement, in contrast to the coarse L0 grid that predicts a slightly diffused shock structure and lower Mach number after the terminal shock. Based on this observation, the L1 grid is deemed adequate for all simulations in this study.

**Table 6 Mesh details for grid convergence study**

| Mesh type | Average element size (mm) | Number of mesh elements (N) | Number of viscous layers |
|---|---|---|---|
| Coarse (L0) | 0.0010 | 53,621 | 22 |
| Medium (L1) | 0.0005 | 201,067 | 44 |
| Fine (L2) | 0.0003 | 542,711 | 74 |

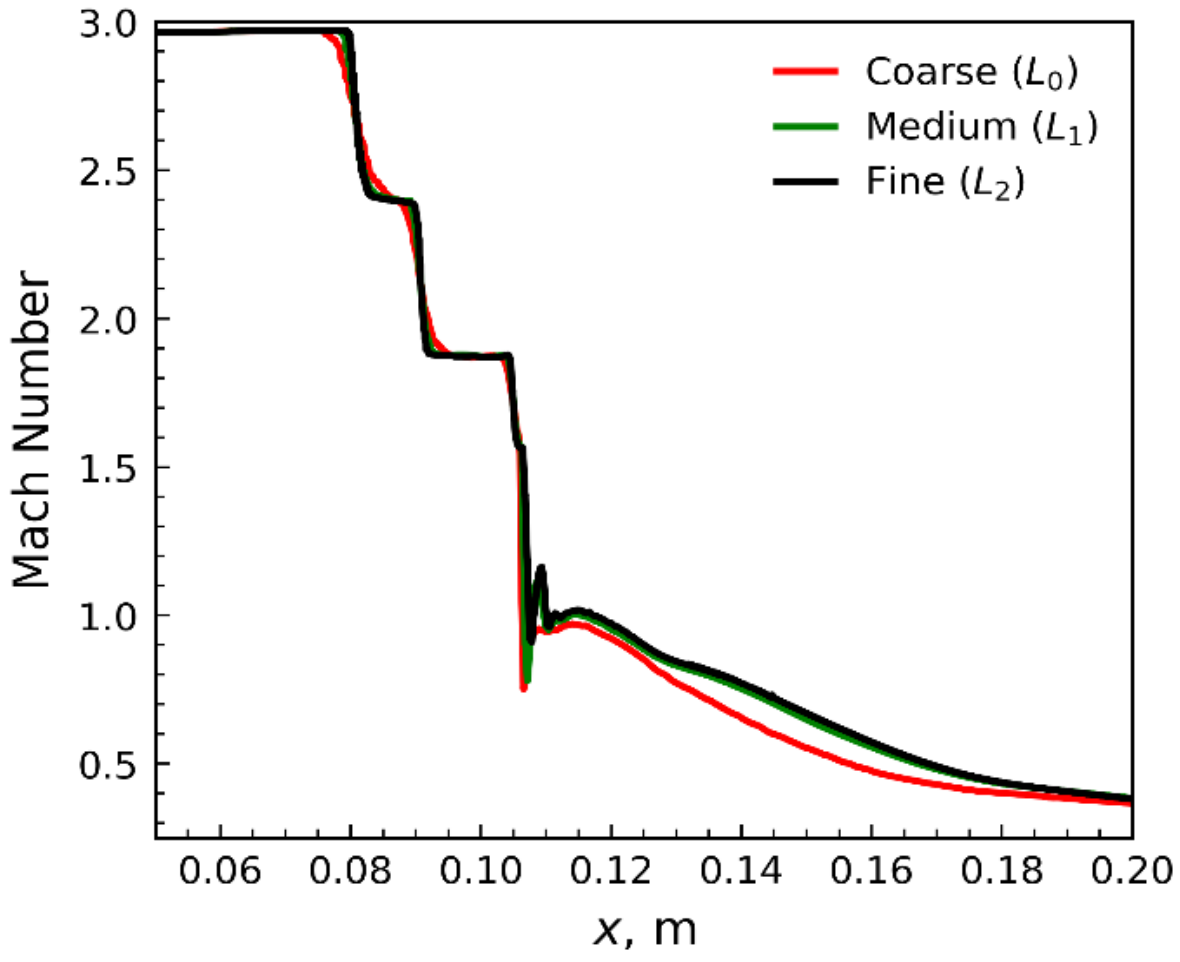

**Fig. 9 Comparison of Mach number along a streamline with different mesh fineness levels.**

Furthermore, to ensure the accuracy of our computational fluid dynamics (CFD) results and establish confidence in our findings, we have conducted an analysis of the observed order of convergence ($p$) [37]. This involves assessing the asymptotic range of convergence for two key performance parameters: the mass-weighted total pressure recovery (TPR) at the exit of the intake and the net drag coefficient due to the intake. These parameters, computed for the three different grid levels, are presented in Table 7. The observed order of convergence is determined to be $p=1.7$ for TPR and $p=1.4$ for the drag coefficient using the code `verify.f90`[7]. Figure 10

[7] Available online at https://www.grc.nasa.gov/www/wind/valid/tutorial/spatconv.html, retrieved Mar. 2025

illustrates this order of convergence with grid size for both functional parameters, with the error calculated using the exact values derived from Richardson Extrapolation. The plot also includes reference dashed lines depicting the first (*p=1*) and second (*p=2*) orders of convergence.

**Table 7 Performance parameters obtained at the three different grid levels used for grid convergence study**

| Mesh type | TPR | $Cd_i$ |
|---|---|---|
| Coarse (L0) | 0.6408 | 0.232 |
| Medium (L1) | 0.6446 | 0.234 |
| Fine (L2) | 0.6462 | 0.235 |

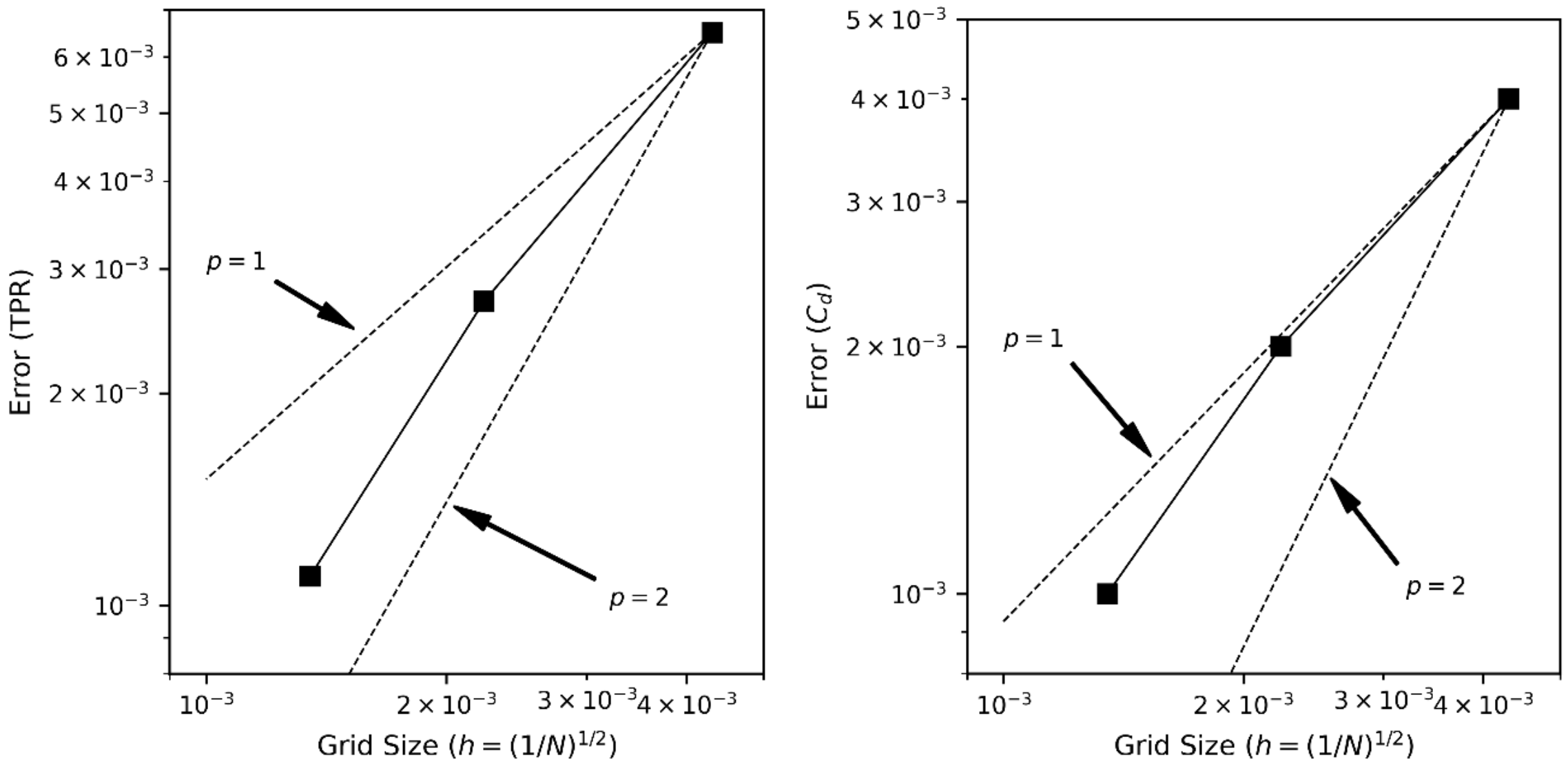

**Fig. 10 Convergence of error in total pressure recovery (left plot) and intake drag coefficient (right plot) with grid size for the three grids in Table 7.**

Additionally, the grid convergence index (GCI) between the medium and fine grids ($\mathrm{GCI}_{12}$) is calculated to be less than 0.5% for both TPR and drag coefficient. A check for asymptotic range of convergence calculated using the expression $\mathrm{GCI}_{12}/r^p \times \mathrm{GCI}_{01}$ yielded a value of approximately 1.5, which is fairly close to 1 and indicates that the solution on the finest grid is in the asymptotic range of convergence. Here, $r$ is the refinement ratio between the grids given by:

$$r = \left(\frac{N_1}{N_2}\right)^{\frac{1}{d}} \tag{3}$$

where $N$ is the total number of grid points used for the corresponding mesh type and $d=2$ is the dimension of the flow domain in this case. As such, all computations for Cases 1 through 6 are performed on the medium (L1) grid size parameters.